\documentclass[aps,prl,showpacs,twocolumn,superscriptaddress]{revtex4-2}
\usepackage{amssymb}
\usepackage{graphicx}
\usepackage{appendix}
\usepackage{dcolumn}
\usepackage{bm}
\usepackage{amsmath}
\usepackage{epstopdf}
\usepackage{xcolor}
\usepackage{float}
\usepackage{braket}
\usepackage[colorlinks,urlcolor=blue,anchorcolor=blue,linkcolor=blue,citecolor=blue,breaklinks=true]{hyperref}
\usepackage[T1]{fontenc}
\usepackage{array}
\usepackage{booktabs}
\begin{document}
\title{Decomposing Fractional Quantum Hall Wave Functions via Operator Contraction Multiplication}
\author{Dong-Hao Guan}
\affiliation{National Laboratory of Solid State Microstructures and Department of Physics, Nanjing University, Nanjing 210093, China}
\author{Licheng Wang}
\affiliation{National Laboratory of Solid State Microstructures and Department of Physics, Nanjing University, Nanjing 210093, China}
\author{Yuan Zhou}
\email{zhouyuan@nju.edu.cn}
\affiliation{National Laboratory of Solid State Microstructures and Department of Physics, Nanjing University, Nanjing 210093, China}
\affiliation{Jiangsu Key Laboratory of Quantum Information Science and Technology, Nanjing University, Suzhou 215163, China}
\author{Ai-Lei He}
\email{heailei@yzu.edu.cn}
\affiliation{College of Physics Science and Technology, Yangzhou University, Yangzhou 225002, China}
\author{Yi-Fei Wang}
\affiliation{Zhejiang Institute of Photoelectronics $\&$ Zhejiang Institute for Advanced Light Source, Zhejiang Normal University, Jinhua 321004, China}
\affiliation{Center for Statistical and Theoretical Condensed Matter Physics, and Department of Physics, Zhejiang Normal University, Jinhua 321004, China}
\date{\today}

\begin{abstract}
We develop a general algebraic scheme to decompose fractional quantum Hall (FQH) wave functions based on the operator contraction multiplication. By introducing fermionic and bosonic operators and establishing three fundamental contraction rules, we achieve an exact decomposition of Laughlin states. This approach naturally extends to multi-component systems by factorizing coupled Jastrow factors via resultants and elementary symmetric polynomials, enabling the first complete decomposition of Halperin states. For Halperin ($2,2,1$) state, we explicitly derive its basic expansion, identify root configurations, and reveal intra- and inter-color squeezing operators, thereby uncovering the underlying generalized Pauli principle. Using this method, we compute orbital entanglement spectra for up to $16$ particles with decomposition dimensions exceeding $10^{11}$, obtaining edge excitation sequences that precisely match chiral Luttinger liquid theory.  Our framework breaks through the longstanding limitations of Jack polynomials, provides a unified decomposition for both single- and multi-component FQH states, and opens a new avenue for exploring wave functions for more complex FQH states.
\end{abstract}
\maketitle

{\noindent {\color{blue} {\it Introduction.}}--- The fractional quantum Hall (FQH) effect~\cite{FQH1,FQH2,FQH3}, as a representative strongly correlated system, has attracted significant interest, particularly regarding the structure and nature of many-body wave functions~\cite{FQH2,WF1,WF2,WF3,WF4,Halperin1,Halperin2,WF5}. Understanding these many-body states requires the lowest Landau level (LLL) orbitals $\phi_m(z)=(2\pi m! 2^m)^{-1/2} z^{m}{\rm exp}(-|z|^2/4)$ (the corresponding angular momentum $L_z=m\hbar$). Such strongly correlated states can be mapped into an effective single-particle orbitals (marked by $z^{m}$ with dropping the Gaussian factor for simplicity)~\cite{Jacks1,Jacks2}, which provides a crucial theoretical paradigm to understand several strongly correlated systems. Within this paradigm, the Jack polynomials (Jacks) offer an efficient decomposition method for both fermionic and bosonic FQH states~\cite{Jacks1,Jacks2,Jacks3,Jacks4,Jacks5,Jacks6,Jacks7,Jacks8,Jacks9}, substantially reducing the computational complexity of the many-body problem. This approach thereby reveals the central role of electron-electron interactions in FQH systems. Remarkably, analysis of the Jacks structure also implies the generalized Pauli principle (GPP) inherent to FQH states~\cite{GPP1,GPP2}, providing the theoretical foundation for fractional statistics.

Over the past twenty years, the Jacks have been proven to be a highly effective method in decomposing several FQH states, such as Laughlin, Pfaffian, and Read-Rezayi states~\cite{Jacks1,Jacks2,Jacks3,Jacks4,Jacks5,Jacks6,Jacks7}. One can easily decompose these FQH states as $\Psi_{\rm FQH}({z_i}) = \sum_{\mu \le \lambda} b_{\mu \le \lambda} m_{\mu}$, where $\lambda$ is the angular momentum partition, $\mu$ is dominated by $\lambda$, $b_{\mu \le \lambda}$ is the expansion coefficients based on the recurrence relation and $m_{\mu}= {\rm Det}(\{z^{\lambda}_i\})$ (${\rm Per}(\{z^{\lambda}_i\})$) is the Slater determinant (permanent) for fermionic (bosonic) systems~\cite{Jacks1,Jacks2,Jacks3,Jacks4,Jacks5}. However, despite these successes, the decomposition of some multi-component FQH states via the Jacks remains challenging~\cite{Jacks6,Jacks7,Jacks9}. Specifically, due to the mathematical complexity inherent in coupled Jastrow factors, current studies have been limited to the multi-component Jacks with special symmetries~\cite{Jacks9}. The existing theoretical frameworks for multi-component Jacks are unable to expand all possible multi-component FQH states and also lack a systematic interpretation of the GPP. A representative example is the Halperin state~\cite{Halperin1,Halperin2} which cannot be decomposed based on the Jacks.

In this work, we propose a universal approach to decompose FQH wave functions based on the contraction multiplication principle, which overcomes the limitations of the Jacks  and provides a unified framework for both single- and multi-component FQH states. By introducing definitions for fermionic and bosonic operators, we establish three fundamental contraction rules to decompose the FQH wave functions. Through the contraction rules, we achieve the exact decomposition of Laughlin states.  More importantly, we further extend this approach to multi-component systems by factorizing coupled Jastrow factors via resultants and symmetric polynomials, enabling the complete decomposition of Halperin states.  Based on the decomposition of FQH wave functions, entanglement spectra are obtained which reveal the topological order of FQH states. This algebraic framework not only yields explicit basis expansions but also reveals the generalized Pauli principle governing Halperin states.

{\noindent {\color{blue} {\it Fermionic and bosonic operators.}}---The Laughlin wave function at $\nu=1/m$ filling in disk geometry is given by~\cite{FQH2},
\begin{equation}\label{Laughlin}
 \begin{aligned}
\Psi_{\rm L}^{\nu=1/m}\left ( \{z_i\} \right) =  \prod_{i<j}{\left( z_i-z_j \right) ^m} = [\prod_{i<j}{\left( z_i-z_j \right)] ^m}
\end{aligned}
\end{equation}
which can be compactly expressed as the $m$th power of a Vandermonde determinant. Here, $\prod_{i<j}{\left( z_i-z_j \right)} = {\rm Det}(\{z^{\lambda}\}) $ is the Vandermonde determinant and $\lambda=[N_{\rm P}-1, N_{\rm P}-2, ..., 1, 0]$ indicates the angular momentum partition with $N_{\rm P}$ particles. $m$ is an odd (even) integer for fermionic (bosonic) states. The Vandermonde determinant encodes the many-particle wave function of noninteracting fermions at integer filling, with its antisymmetric structure emerging naturally from the Slater determinant of the occupied single-particle states. Here, we define the fermionic (antisymmetric) operator $\bf F$, which originates from the Vandermonde determinant, i.e., ${\rm Det}(\{z^{\lambda}\})={\bf F}_{\lambda}$. The $\nu=1/m$ Laughlin state is expressed as $\Psi _{\rm L} ^{\nu=1/m}\left ( \{z\} \right) = ({\bf F}_{\lambda})^m$. Similarly, one can define the bosonic (symmetric) operator $\bf B$ which is the permanent, i.e., ${\rm Per}(\{z^{\lambda}\}) = {\bf B}_{\lambda}$, corresponding to the wave function for free bosons. Differing from the bosonic operator ${\bf B}_{\lambda}$, all elements in $\lambda$ must be different for the fermionic operator ${\bf F}_{\lambda}$, due to the Pauli principle.

To illustrate the newly defined operators $\bf F$ and $\bf B$ more clearly, we respectively consider wave functions with three free fermions and bosons as concrete examples (i.e., $\lambda=[2, 1, 0]$). For noninteracting fermions, the wave function can be expanded as,
\begin{equation}\label{VD1}
 \begin{aligned}
{\rm Det}(\{z^{\lambda}\}) = \sum_{\sigma(\lambda)} \tilde{\varepsilon}_{\sigma(\lambda)}z^{\lambda_1}_1z^{\lambda_2}_2z^{\lambda_3}_3 \equiv{ {\bf F}_{[210]}}.
\end{aligned}
\end{equation}
Here, $\tilde{\varepsilon}_{\sigma(\lambda)} = (-1)^{\lfloor \frac{{ N_{\rm P}}}{2} \rfloor} \varepsilon_{\sigma(\lambda)}$ is the signed Levi-Civita symbol, with ${\varepsilon}_{\sigma(\lambda)}$ taking values $+1$ or $-1$ according to the permutation parity, and $\lfloor \dots \rfloor$ is the floor function, and ${\sigma(\lambda)}$ indicates all permutations of $\lambda$. $\lambda_i$ is the $i$th element of $\lambda$. For free bosons, the wave function can be expanded as,
 \begin{equation}\label{VD2}
 \begin{aligned}
 {\rm Per}(z^{\lambda}_1,z^{\lambda}_2,z^{\lambda}_3)  =  \sum_{\sigma(\lambda)}  z^{\lambda_1}_1z^{\lambda_2}_2z^{\lambda_3}_3 \equiv{ {\bf B}_{[210]}}.
\end{aligned}
\end{equation}
Unlike the fermionic operator, bosonic operator possesses exchange symmetry and allows multiple bosons occupying in the same orbital. When considering three free bosons loaded into the orbitals $\lambda=[4,1,1]$, one can easily expand  ${\rm Per}(z^{\lambda}_1,z^{\lambda}_2,z^{\lambda}_3) = 2(z^4_1z^1_2z^1_3+z^1_1z^4_2z^1_3+z^1_1z^1_2z^4_3)=2{\bf B}_{[411]}$. Thus, the bosonic operator is redefined as ${\rm Per}(\{z\})= \#_{\lambda} {\bf B}_{\lambda}$, where $\#$ counts orbital multiplicities, $\#_{\lambda}=\prod_{\lambda_i}(N_{\lambda_i})!$, and $N_{\lambda_i}$ is the number of particles occupying in the orbital with angular momentum $\lambda_i$. For $\lambda=[4,1,1]$, there are two bosons occupy the $m=1$ orbital, so $\#_{[4,1,1]}=2!=2$.

Starting from the fundamental properties of fermionic and bosonic operators, we can derive their operational rules:\\ \textit{
i) The product of an even number of fermionic operators yields a bosonic operator;\\
ii) The product of an odd number of fermionic operators remains a fermionic operator;\\
iii) The product of a fermionic operator with a bosonic operator is fermionic;\\
iv) The product of two bosonic operators is bosonic.}\\
Accordingly, for any pair of operators, we can perform three types of contraction operations: fermion-fermion (${\rm F-F}$), fermion-boson (${\rm F-B}$), and boson-boson (${\rm B-B}$).

{\noindent {\color{blue} {\it Operator contraction multiplication.}}---We define rules of these three contraction multiplication:
\begin{eqnarray}
  \bf{F}_{\lambda}{\bf F}_{\lambda^\prime} &=& \sum_{\sigma \left( \lambda^\prime \right)}{\tilde{\varepsilon}_{\sigma \left( \lambda^\prime \right)}\#_{\lambda +\sigma \left( \lambda^\prime \right)} {\bf B}_{[{\lambda +\sigma \left( \lambda^\prime \right)}]^\prime}}, \label{ff}\\
  {\bf B}_{\lambda} {\bf B}_{\lambda^\prime} &=& \frac{1}{\#_{\lambda}}\sum_{\tilde{\sigma}\left( \lambda^\prime \right)}{\#_{\lambda +\tilde{\sigma}\left( \lambda^\prime \right)} {\bf B}_{\lambda +\tilde{\sigma}\left( \lambda^\prime \right)}}, \label{bb}\\
  {\bf F}_{\lambda} {\bf B}_{\lambda^\prime} &=& \sum_{\tilde{\sigma}\left( \lambda^\prime \right)} {\tilde{\varepsilon}_{\lambda +\tilde{\sigma}\left( \lambda^\prime \right)} {\bf F}_{\left[ \lambda +\tilde{\sigma}\left( \lambda^\prime \right) \right]^{ \prime}}}. \label{fb}
\end{eqnarray}
Here, $\sigma \left( \lambda \right)$ is all permutation of $\lambda$. The reduced permutation, denoted by $\tilde{\sigma}\left( \lambda \right)$, accounts for the multiple occupancy.  $\left[ \lambda +\tilde{\sigma}\left( \lambda^\prime \right) \right]^{ \prime}$ is the partition sorted in descending order. One can see that Eqs.~(\ref{ff})-(\ref{fb}) are of the ${\rm F-F}$, ${\rm B-B}$, and ${\rm F-B}$ types, respectively.

\begin{figure}[h]
    \centering
    \includegraphics[width=\columnwidth]{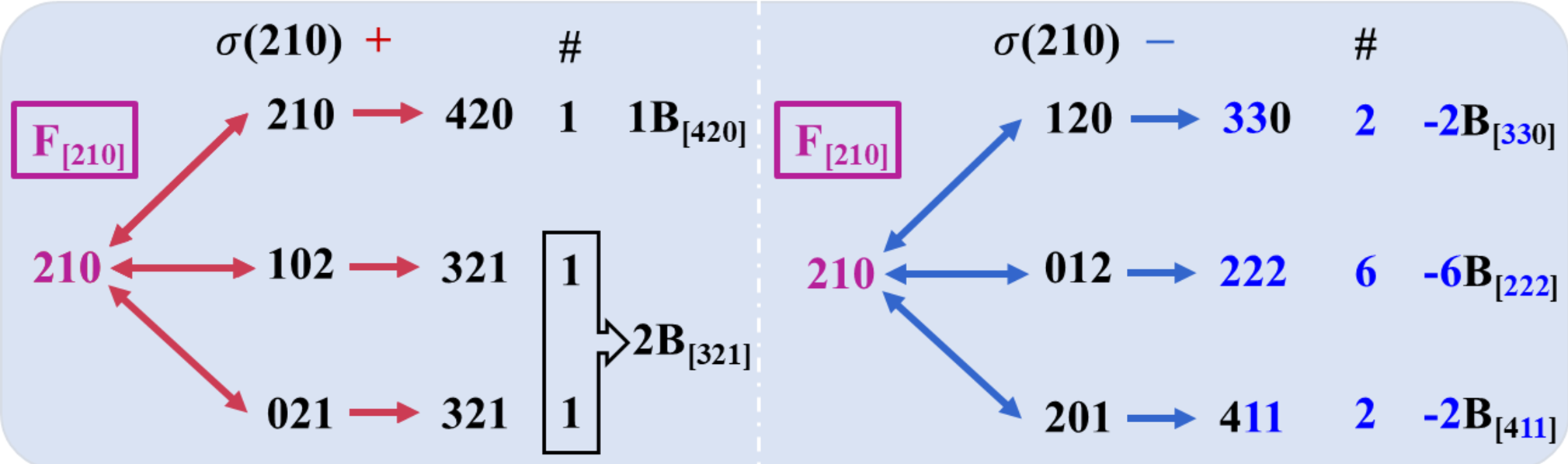}
    \caption{Schematic Diagram of decomposing $\nu=1/2$ Laughlin state with three particles based on the contraction rule in ${\rm F-F}$ type. We enumerate all permutations of $[2, 1, 0]$ as $\sigma(210)$ and add them termwise to $\lambda = [2, 1, 0]$. We also calculate the corresponding permutation signs($\tilde{\varepsilon}$) and bosonic multiple occupation coefficients($\#$). Summing over all these terms yields the decomposition of $\nu = 1/2$ Laughlin state.}
    \label{fig1}
\end{figure}

Laughlin states can be decomposed based on contraction rules. For simplicity, we take a $\nu=1/2$ Laughlin state with three bosons as an example, the corresponding wave function is $\Psi _{\rm L}^{\nu=1/2}\left ( z_1, z_2, z_3 \right) = ( {\bf F}_{[210]})^2 = {\bf F}_{\lambda}{\bf F}_{\lambda^{\prime}}$ which belongs to ${\rm F-F}$ type with $\lambda=\lambda^{\prime}=[2,1,0]$. One can easily obtain $\sigma([2,1,0]) = \{[2,1,0],[1,0,2],[0,2,1], [1,2,0], [0,1,2], [2,0,1]\}$ (as shown in Fig.~\ref{fig1}).  $\tilde{\varepsilon}\left( [2,1,0] \right) = \{1,\ 1,\ 1,\ -1,\ -1,\ -1\}$ and $\#_{[2,1,0]+\sigma([2,1,0])}$ are $\{1,\ 1,\ 1,\ 2,\ 6,\ 2\}$. According to Eq.~(\ref{ff}), $\nu=1/2$ Laughlin state with three bosons can be decomposed as $\Psi _{\rm L}^{\nu=1/2}\left ( z_1, z_2, z_3 \right) = {\bf B}_{[420]} + 2{\bf B}_{[321]} -2{\bf B}_{[411]}  -2{\bf B}_{[330]}  -6{\bf B}_{[222]}$ (shown in Fig.~\ref{fig1}). $\nu=1/4$ ($\nu=1/3$) Laughlin state belongs to the ${\rm B-B}$ (${\rm F-B}$) type, one can easily expand these Laughlin wave functions based on the contraction rules [details shown in the Supplemental Material (SM)~\cite{SM}].

\begin{figure*}[ht]
    \centering
    \includegraphics[width=\textwidth]{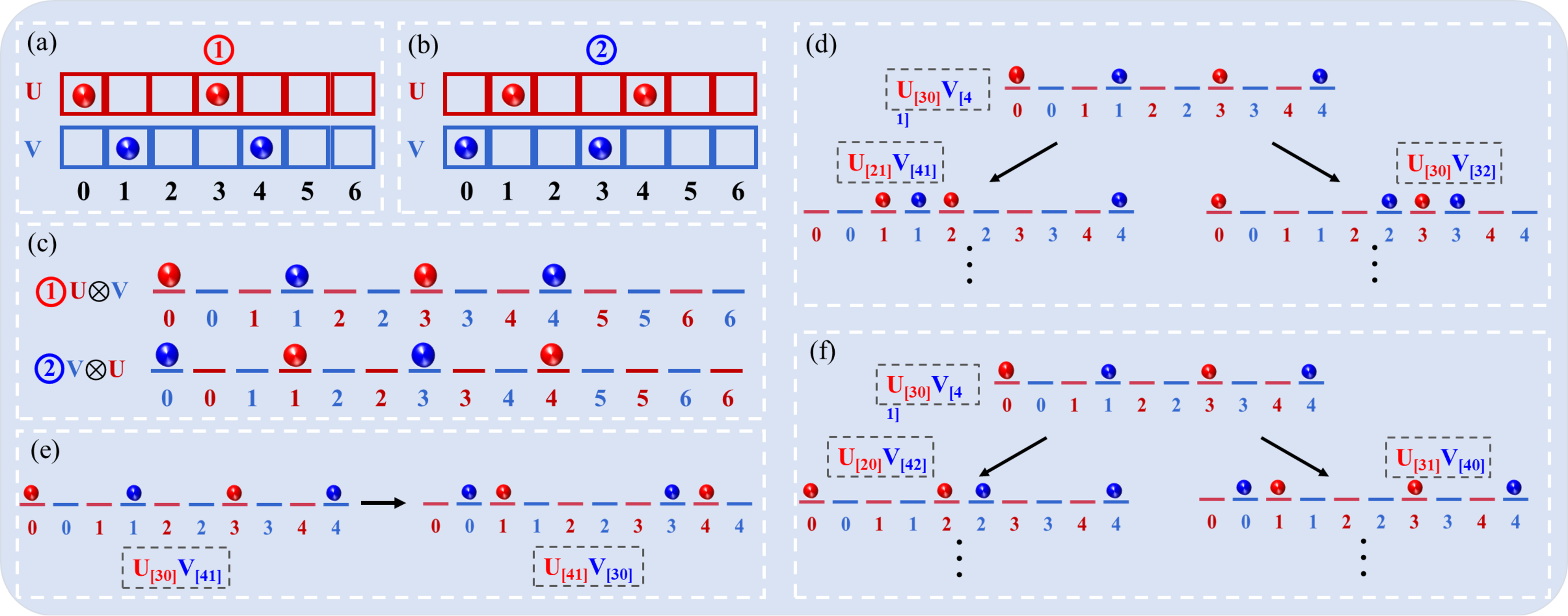}
    \caption{Root configurations of particle occupation and parts of squeezing operation in occupation language from the root configuration for Halperin $(2,2,1)$ state with four particles filling in the ``entangled'' orbitals. ``U''  and ``V''  denote the degree of freedom (such as spin, layer, etc) colored with red and blue, respectively. Two root configurations are presented in (a) and (b). (c). ``U'' and ``V'' orbitals can be entangled each other and lead to ``entangled orbtials'', which can be expressed as ${\bf U} \otimes {\bf V}$ or ${\bf V} \otimes {\bf U}$. Particles squeezing occurs in (d) intra-color orbitals and (e),(f) inter-color orbitals. (e) ${\bf U}_{[41]} {\bf V}_{[30]}$ can be obtained by squeezing particles of ${\bf U}_{[30]} {\bf V}_{[41]}$. Colors of the orbitals indicate various degree of freedom.  Numbers below the orbitals indicate the angular momenta.}
    \label{fig2}
\end{figure*}

{\noindent {\color{blue} {\it Coupled Jastrow factors.}}---The decomposition of the multi-component FQH states (such as Halperin states) remains challenging due to their multi-component nature and the intrinsically coupled Jastrow factors. For clarity, we begin to consider the two-component $(m,m,n)$ Halperin wave function~\cite{Halperin1,Halperin2}, i.e.,
\begin{equation}\label{halperin}
\Psi_{\rm H}^{(mmn)} =\prod_{i<j}{\left( u_i-u_j \right) ^m}\prod_{i<j}{\left( v_i-v_j \right) ^m}\prod_{i,j}{\left( u_i-v_j \right) ^n},
\end{equation}
where $u_i$ and $v_i$ are the complex coordinates of the $i$th electron with different degree of freedom (such as spin, layer, etc).
We have completed the expansion of the Laughlin state following the operator contraction rules, and the first two product terms in the Halperin wave function [Eq.~(\ref{halperin})] are the $\nu=1/m$ Laughlin wave function which can be straightforwardly decomposed. The primary difficulty is to expand the coupled Jastrow factor [the third term in Eq.~(\ref{halperin})].

We rewrite the coupled Jastrow factor,
\begin{equation}\label{CJF}
{\cal J}_{(N_u,N_v)}^n = J_{(N_u,N_v)} (\{u_i,v_j\}) = \prod^{N_u}_{i=1} \prod^{N_v}_{j=1} {\left( u_i-v_j \right) ^n},
\end{equation}
where $N_u$ ($N_v$) denotes the particle numbers of the $u$ ($v$) component. One can find that the coupled Jastrow term equals the resultant of two polynomials ${\cal P}(z)$ and ${\cal Q}(z)$ whose roots are $\{ u_i \}$ and $\{ v_j \}$, and the Jastrow factor ${\cal {J}}_{(N_u,N_v)} = {\rm Res} ({\cal P}, {\cal Q})$, where ${\rm Res} ({\cal P}, {\cal Q})$ is the resultant of $\cal P$ and $\cal Q$. The polynomials can be easily constructed as,
\begin{equation}\label{Polys}
{\cal P} \left( z \right) =\prod_{i=1}^{N_u} {\left( z-u_i \right)},\quad  {\cal Q} \left( z \right) =\prod_{j=1}^{N_v} {\left( z-v_j \right)},
\end{equation}
and $\{u_i\}$ and $\{v_j\}$ are the roots of ${\cal P}(z)=0$ and ${\cal Q}(z)=0$, respectively.

According to Viete theorem, ${\cal P} (z)$ and ${\cal Q} (z)$ can be expanded as,
\begin{eqnarray*}\label{viete1}
\mathrm{\cal P}\left( z \right) &=& z^{N_u}+\tilde{U}_1z^{N_u-1}+\cdots +\tilde{U}_kz^{N_u-k}+\cdots+\tilde{U}_N, \\
\mathrm{\cal Q}\left( z \right) &=& z^{N_v}+\tilde{V}_1z^{N_v-1}+\cdots +\tilde{V}_kz^{N_v-k}+\cdots+\tilde{V}_N.
\end{eqnarray*}
Here, $\tilde{U}_k$ ($\tilde{V}_k$) is the $k$-th elementary symmetric polynomial with respect to $u$ ($v$), and
\begin{equation}\label{viete2}
 \tilde{U}_k = (-1)^k \sum_{1 \le i_1 < i_2  < \cdots <i_k \le N_u} u_{i_1} u_{i_2} \cdots u_{i_k}.
\end{equation}
Inspired by Eq.~(\ref{VD2}), $\tilde{U}_k$ and $\tilde{V}_k$ can be defined as the bosonic operators ${\bf \tilde{ U}}_{\lambda_k}$ and ${\bf \tilde{ V}}_{\lambda_k}$. The angular momentum partitions are $\lambda_k = [1^k0^{N_u-k}]$ for ${\bf \tilde{ U}}_{\lambda_k}$ and $\lambda_k = [1^k0^{N_v-k}]$ for ${\bf \tilde{ V}}_{\lambda_k}$, where ``$1^k$'' (or ``$0^{N-k}$'') means a sequence of $k$ (or $N-k$) particles occupying the ``orbitals'' with angular momentum $1$ (or $0$).
The resultant of ${\cal P}(z)$ and ${\cal Q}(z)$, i.e., ${\cal J}_{(N_u,N_v)} = \mathrm{Res(} {\cal P}, {\cal Q})$, can be calculated by Sylvester matrix. When considering $N_u=N_v=N$, the Sylvester matrix is,
\begin{equation}\label{Resx}
 \begin{aligned}
{\cal J} =\left\vert \begin{matrix}
	1&		0&		\cdots&		0&		1&		0&		\cdots&		0\\
	{\bf \tilde{ U}}_{\lambda_1}&        1&		\cdots&		0&		{\bf \tilde{V}}_{\lambda_1}&		1&		\cdots&		0\\
	{\bf \tilde{U}}_{\lambda_2}&	  {\bf \tilde{U}}_{\lambda_1}&		\cdots&		0&		{\bf \tilde{V}}_{\lambda_2}&		{\bf \tilde{V}}_{\lambda_1}&		\cdots&		0\\
	\vdots&		\vdots&		\ddots&		\vdots&		\vdots&		\vdots&		\ddots&		\vdots\\
	{\bf \tilde{U}}_{\lambda_{N}}&	{\bf \tilde{U}}_{\lambda_{N-1}}&		\cdots&		1&		{\bf \tilde{V}}_{\lambda_N}&		{\bf \tilde{V}}_{\lambda_{N-1}}&		\cdots&		1\\
	0&		{\bf \tilde{U}}_{\lambda_{N}}&		\cdots&		{\bf\tilde{ U}}_{\lambda_1}&		0&		{\bf \tilde{V}}_{\lambda_N}&		\cdots&		{\bf \tilde{ V}}_{\lambda_1}\\
	\vdots&		\vdots&		\ddots&		\vdots&		\vdots&		\vdots&		\ddots&		\vdots\\
	0&		0&		\cdots&		{\bf \tilde{V}}_{\lambda_N}&		0&		0&		\cdots&		{\bf \tilde{V}}_{\lambda_N}\\
\end{matrix} \right\vert.
\end{aligned}
\end{equation}
Here, ${\bf \tilde{U}}_{\lambda_k}=(-1)^k {\bf U}_{\lambda_k}$, and ${\bf \tilde{V}}_{\lambda_k}=(-1)^k {\bf V}_{\lambda_k}$.
Two key rules of the operator multiplication for the multi-component FQH states are, \\
(i) \emph {Multiplication between symmetric operators with distinct names (such as ${\bf U}_{\lambda} - {\bf V}_{\lambda}$) follows a direct product}; \\
(ii) \emph {Operators sharing the same name (such as ${\bf U}_{\lambda} - {\bf U}_{\lambda^\prime}$ or ${\bf V}_{\lambda} - {\bf V}_{\lambda^\prime}$) are contracted}.

{\noindent {\color{blue} {\it Decomposing Halperin states.}}---We illustrate the contraction rule concretely using the four-particle Halperin $(2,2,1)$ state as an example, and the corresponding wave function is,
\begin{equation}\label{Halpx}
 \begin{aligned}
\Psi_{\rm H}^{(221)} &=\prod_{i<j}{\left( u_i-u_j \right) ^2}\prod_{i<j}{\left( v_i-v_j \right) ^2}\prod_{i,j}{\left( u_i-v_j \right)}\\
&=\left( u_1-u_2 \right) ^2\left( v_1-v_2 \right) ^2 \left( u_1-v_1 \right) \left( u_2-v_1 \right) \\
&\quad\times  \left( u_1-v_2 \right) \left( u_2-v_2 \right) ={{\bf U}_{10}^{\prime}}^2  {{\bf V}_{10}^{\prime}}^2 \mathcal{J}_{(2,2)} .
\end{aligned}
\end{equation}
For the multi-component states, the fermionic and bosonic operators are respectively denoted by ${\bf U}^{\prime}$ (${\bf V}^{\prime}$) and ${\bf U}$ (${\bf V}$), instead of $\bf F$ and $\bf B$. According to the ${\bf F}-{\bf F}$ contraction rule [see Eq.~(\ref{ff})],
\begin{equation}\label{halpx1}
 \begin{aligned}
{{\bf U}_{[10]}^{\prime}}^2 = {\bf U}_{[20]}-2 {\bf U}_{[11]},\  {{\bf V}_{[10]}^{\prime}}^2={\bf V}_{[20]}-2{\bf V}_{[11]}.
\end{aligned}
\end{equation}

According to Eq.~(\ref{Resx}), one can easily obtain ${\cal J}_{(2,2)}$, i.e.,
\begin{equation}\label{Jas22}
 \begin{aligned}
\mathcal{J}_{( 2,2)} &=\left| \begin{matrix}
	1&		                  0&		1&		0\\
	-{\bf U}_{[10]}&		  1&		-{\bf V}_{[10]}&	1	\\
	{\bf U}_{[11]}&		-{\bf U}_{[10]}&		{\bf V}_{[11]}&		-{\bf V}_{[10]}\\
	0&		                 {\bf U}_{[11]}&		0&		{\bf V}_{[11]}\\
\end{matrix} \right|
=\left( {\bf U}_{[11]} \right) ^2+
\\&\left( {\bf V}_{[11]} \right) ^2 -{\bf V}_{[10]} {\bf U}_{[10]} {\bf U}_{[11]}- {\bf U}_{[10]} {\bf V}_{[10]} {\bf V}_{[11]} + \\
& {\bf V}_{[11]} ( {\bf U}_{[10]}) ^2 +  {\bf U}_{[11]}\left( {\bf V}_{[10]} \right) ^2-2{\bf U}_{[11]}{\bf V}_{[11]}.
\end{aligned}
\end{equation}
Based on the ${\bf B}-{\bf B}$ contraction rule, the coupled Jastrow factor is,
\begin{equation}\label{Jas221}
 \begin{aligned}
&\mathcal{J}_{( 2,2)} = {\bf U}_{[22]}+{\bf V}_{[22]}-{\bf V}_{[10]}{\bf U}_{[21]}-{\bf U}_{[10]}{\bf V}_{[21]} \\
& +{\bf V}_{[11]}{\bf U}_{[20]}+{\bf V}_{[20]}{\bf U}_{[11]}+2{\bf V}_{[11]}{\bf U}_{[11]}.
\end{aligned}
\end{equation}
Combining Eqs.~(\ref{halpx1}) and ~(\ref{Jas221}), the Halperin $(2,2,1)$ wave function can be factorized as,
\begin{equation}\label{halpx2}
 \begin{aligned}
\Psi_{\rm H}^{\left(221 \right)}&=\mathbf{U}_{\left[ 10 \right]}^{\prime}\mathbf{V}_{\left[ 10 \right]}^{\prime}\mathcal{J} _{\left( 2,2 \right)}= (\mathbf{U}_{\left[ 42 \right]}\mathbf{V}_{\left[ 20 \right]}-2\mathbf{U}_{\left[ 42 \right]}\mathbf{V}_{\left[ 11 \right]}
\\
&+4\mathbf{U}_{\left[ 33 \right]}\mathbf{V}_{\left[ 11 \right]}-2\mathbf{U}_{\left[ 33 \right]}\mathbf{V}_{\left[ 20 \right]}-\mathbf{U}_{\left[ 41 \right]}\mathbf{V}_{\left[ 30 \right]}
\\
&+\mathbf{U}_{\left[ 41 \right]}\mathbf{V}_{\left[ 21 \right]}-\mathbf{U}_{\left[ 32 \right]}\mathbf{V}_{\left[ 21 \right]}+\mathbf{U}_{\left[ 32 \right]}\mathbf{V}_{\left[ 30 \right]}
\\
&-2\mathbf{U}_{\left[ 40 \right]}\mathbf{V}_{\left[ 22 \right]}+2\mathbf{U}_{\left[ 31 \right]}\mathbf{V}_{\left[ 22 \right]}+\mathbf{U}_{\left[ 40 \right]}\mathbf{V}_{\left[ 31 \right]}
\\
&-\mathbf{U}_{\left[ 31 \right]}\mathbf{V}_{\left[ 31 \right]})+(\mathbf{U}\leftrightarrow \mathbf{V}).
\end{aligned}
\end{equation}
Here, $\bf{U}\leftrightarrow \bf{V}$ indicates that $\bf U$ and $\bf V$ are swapped.

The expansion form of the Halperin $(2,2,1)$ state [Eq.~(\ref{halpx2})] allows for the derivation of its corresponding root configurations, i.e., ${\bf U}_{[30]} {\bf V}_{[41]}$ or ${\bf U}_{[41]} {\bf V}_{[30]}$. To better illustrate the root configuration of Halperin $(2,2,1)$ state with four particles, a schematic of the particle occupation is shown in Fig.~\ref{fig2} (a) and (b), and orbitals are color-coded according to their degree of freedom [i.e., $\bf U$ ($\bf V$) orbitals colored with red (blue)]. The orbitals can be ``entangled'' together and particles occupied the ``entangled'' orbitals following the pervious occupation [as shown in Fig.~\ref{fig2} (c)]. One can easily find the Halperin $(2,2,1)$ state with $\nu = 1/3$ filling in the ``entangled'' orbitals. The root configuration seems to be ``1001001001'' in occupation language based on the ``entangled'' orbitals, and it also obeys the squeezing rule.

\begin{figure}[htp]
\includegraphics[width=\columnwidth]{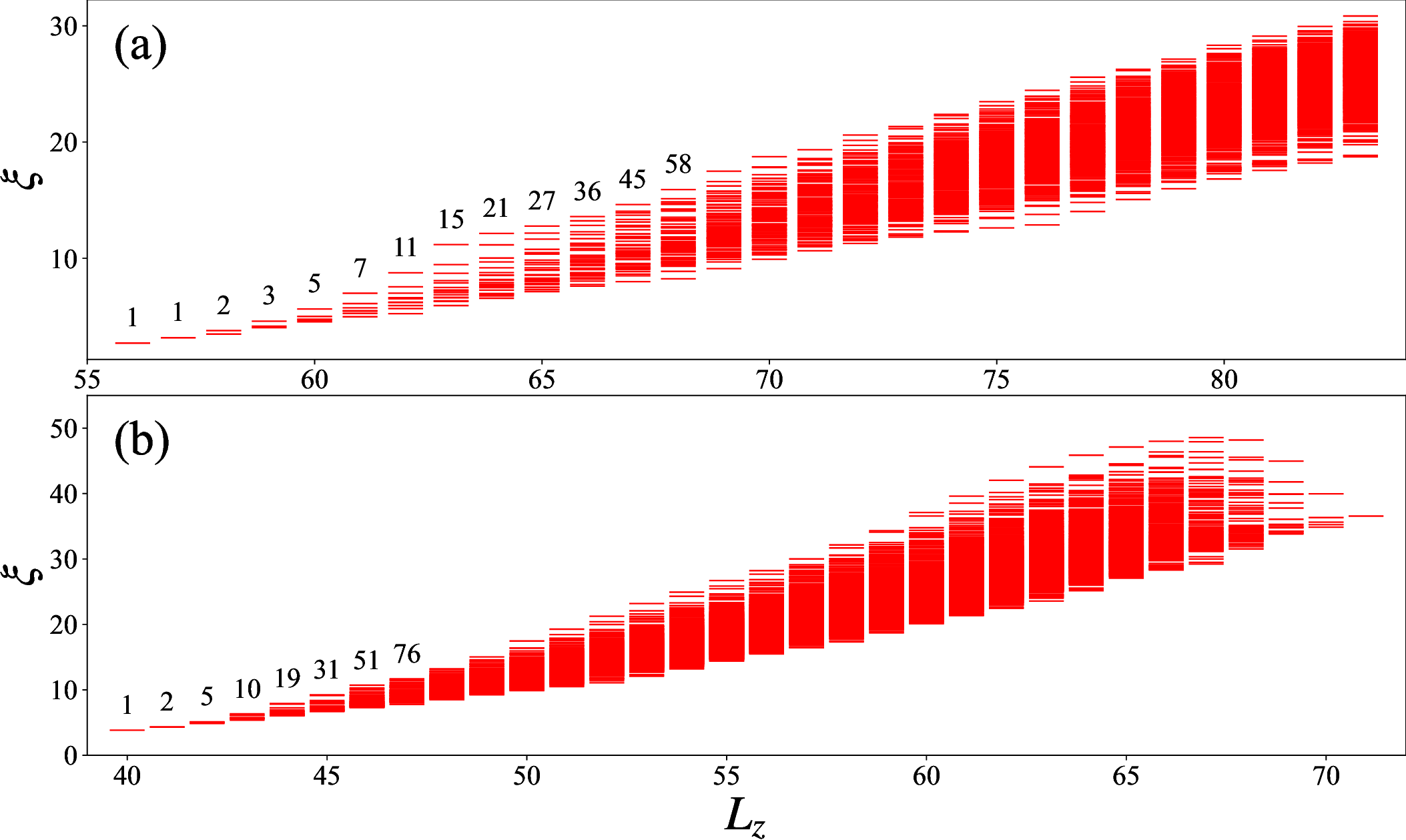}
\caption{(Color online) Orbital entanglement spectra for (a) $\nu=1/2$ Laughlin and (b) Halperin $(2,2,1)$ states with 16 particles in disk geometry. Here, each system is partitioned into two subsystems $\rm A$ and $\rm B$. The bipartition parameters for (a) $\nu=1/2$ Laughlin state are $N^{\rm A}=8$ and $L^{\rm A}=16$, and (b) $\nu=1/3$ Halperin state are $N_{u}^{\rm A}=4$, $N_{v}^{\rm A}=4$, $L_{u}^{\rm A}=12$ and $L_{v}^{\rm A}=12$. $N^{\rm A}$ is the number of particles in $\rm A$ subsystem, and two subspaces are partitioned at angular momentum $L^{\rm A}$. ``{$u$}'' (``{$v$}'') indicates various orbital degree of freedom.}
\label{fig3}
\end{figure}

We illustrate the squeezing operation for the Halperin $(2,2,1)$ state with four particles in Fig.~\ref{fig2}. The corresponding root configuration is ${\bf U}_{[30]} {\bf V}_{[41]}$ and other basis can be obtained based on the root configuration by squeezing operation~\cite{Jacks1,Jacks2,Jacks3,Jacks4,Jacks5}. However, differing from the Jacks states, there are two types of squeezing operations for the Halperin $(2,2,1)$ state, i.e., particle squeezing in intra-color orbitals  and inter-color orbitals [details in Fig.~\ref{fig2} (d)-(f)]. When squeezing occurs between particles in intra-color orbitals, the angular momentum of the same-color orbitals is conserved [as illustrated in Fig.~\ref{fig2} (d)]. Squeezing between inter-color orbitals leads to the breakdown of angular momentum conservation for the same-color orbitals, such as ${\bf U}_{[20]}{\bf V}_{[42]}$, ${\bf U}_{[31]}{\bf V}_{[40]}$ and ${\bf U}_{[40]}{\bf V}_{[31]}$ [shown in Figs.~\ref{fig2} (e) and (f)]. Additionally, configuration ${\bf U}_{[41]} {\bf V}_{[30]}$ can be obtained by squeezing particles in the inter-color orbitals twice based on the root configuration ${\bf U}_{[30]} {\bf V}_{[41]}$ [in Fig.~\ref{fig2} (e)]. ${\bf U}_{[30]} {\bf V}_{[41]}$ can be obtained by squeezing operation of ${\bf U}_{[41]} {\bf V}_{[30]}$ as well. Accordingly, we can choose either configuration as the root configuration. And other basis configurations can be obtained by the squeezing operation based on the root configuration (details in the SM~\cite{SM}). Although our decomposing method does not involve particle squeezing operations, such a squeezing process can provide a clear physical picture for understanding Halperin states and further demonstrates their consistency with the GPP. Likewise, other Halperin states admit an analogous decomposition (for example, the decomposition of Halperin (3,3,2) state in the SM~\cite{SM}).

Using this decomposition method, the $\nu=1/2$ Laughlin and Halperin $(2,2,1)$ states are decomposed for 16 particles, with decomposition dimensions  up to $10^9$ and $10^{11}$, respectively (details of basis dimensions in the SM~\cite{SM}). Topological order in FQH states is characterized by entanglement properties, via the correspondence between entanglement spectra and edge excitations~\cite{LiHaldane}. Based on the decomposition of the $\nu=1/2$ Laughlin and Halperin $(2,2,1)$ states, their orbital entanglement spectra in disk geometry are presented in Fig.~\ref{fig3} (details of the calculation given in the SM~\cite{SM}). Notably, edge excitations for these two FQH states are obtained with the sequences ``1, 1, 2, 3, 5, 7,...'' and ``1, 2, 5, 10,...'', which are consistent with the predictions of the chiral Luttinger liquid theory~\cite{XGWen}. These entanglement spectra reveal different topological orders of various FQH states. Moreover, density profile of Laughlin and Halperin states can be obtained as well (details in the SM~\cite{SM}).

{\noindent {\color{blue} {\it Summary.}}---We develop a systematic operator contraction multiplication framework for decomposing FQH wave functions. By introducing fermionic and bosonic operators and establishing three fundamental contraction rules, we achieve exact decomposition of Laughlin states into symmetric or antisymmetric bases. This approach is further extended to multi-component FQH by factorizing coupled Jastrow factors via resultants and elementary symmetric polynomials, enabling the first complete decomposition of Halperin states. In addition, this decomposition yields the entanglement spectrum of various FQH states.

Our work provides a unified algebraic framework that not only yields explicit basis expansions for FQH states but also reveals the underlying GPP governing multi-component systems. This decomposition method enables the construction of trial wave functions for fractional Chern insulators (FCIs). Although several wave functions for FCIs have been constructed based on the Jacks and GPP~\cite{He_2015,NAFCIs1}, the construction remains challenging for color-entangled FCI states~\cite{colorent1,colorent2} and multi-component FQH states in topological flat band models~\cite{zeng1,zeng2,zeng21,zeng3}. For various novel FCI states recently observed in moir{\'e} super-lattice materials~\cite{ReaFCI1,ReaFCI2,ReaFCI3,ReaFCI4,ReaFCI5,ReaFCI6,ReaFCI7}, their nature remains unclear. Trial wave functions can thus effectively capture the essential physics of these states, including their corresponding GPP.

{ {\it Acknowledgments}}---This work was supported by the NSFC under Grants No. 11874325, No. 12204404 and No. 12374137, Quantum Science and Technology-National Science and Technology Major Project Grant No.2025ZD0300400, and the Natural Science Foundation of Jiangsu Province Grant No. BK20231397.

{ {\it Data availability}}---The data that support the findings of this article are not publicly available. The data are available from the authors upon reasonable request.

\bibliography{FQH}

\begin{thebibliography}{39}%
\makeatletter
\providecommand \@ifxundefined [1]{%
 \@ifx{#1\undefined}
}%
\providecommand \@ifnum [1]{%
 \ifnum #1\expandafter \@firstoftwo
 \else \expandafter \@secondoftwo
 \fi
}%
\providecommand \@ifx [1]{%
 \ifx #1\expandafter \@firstoftwo
 \else \expandafter \@secondoftwo
 \fi
}%
\providecommand \natexlab [1]{#1}%
\providecommand \enquote  [1]{``#1''}%
\providecommand \bibnamefont  [1]{#1}%
\providecommand \bibfnamefont [1]{#1}%
\providecommand \citenamefont [1]{#1}%
\providecommand \href@noop [0]{\@secondoftwo}%
\providecommand \href [0]{\begingroup \@sanitize@url \@href}%
\providecommand \@href[1]{\@@startlink{#1}\@@href}%
\providecommand \@@href[1]{\endgroup#1\@@endlink}%
\providecommand \@sanitize@url [0]{\catcode `\\12\catcode `\$12\catcode `\&12\catcode `\#12\catcode `\^12\catcode `\_12\catcode `\%12\relax}%
\providecommand \@@startlink[1]{}%
\providecommand \@@endlink[0]{}%
\providecommand \url  [0]{\begingroup\@sanitize@url \@url }%
\providecommand \@url [1]{\endgroup\@href {#1}{\urlprefix }}%
\providecommand \urlprefix  [0]{URL }%
\providecommand \Eprint [0]{\href }%
\providecommand \doibase [0]{https://doi.org/}%
\providecommand \selectlanguage [0]{\@gobble}%
\providecommand \bibinfo  [0]{\@secondoftwo}%
\providecommand \bibfield  [0]{\@secondoftwo}%
\providecommand \translation [1]{[#1]}%
\providecommand \BibitemOpen [0]{}%
\providecommand \bibitemStop [0]{}%
\providecommand \bibitemNoStop [0]{.\EOS\space}%
\providecommand \EOS [0]{\spacefactor3000\relax}%
\providecommand \BibitemShut  [1]{\csname bibitem#1\endcsname}%
\let\auto@bib@innerbib\@empty
\bibitem [{\citenamefont {Tsui}\ \emph {et~al.}(1982)\citenamefont {Tsui}, \citenamefont {Stormer},\ and\ \citenamefont {Gossard}}]{FQH1}%
  \BibitemOpen
  \bibfield  {author} {\bibinfo {author} {\bibfnamefont {D.~C.}\ \bibnamefont {Tsui}}, \bibinfo {author} {\bibfnamefont {H.~L.}\ \bibnamefont {Stormer}},\ and\ \bibinfo {author} {\bibfnamefont {A.~C.}\ \bibnamefont {Gossard}},\ }\bibfield  {title} {\bibinfo {title} {Two-dimensional magnetotransport in the extreme quantum limit},\ }\href {https://doi.org/10.1103/PhysRevLett.48.1559} {\bibfield  {journal} {\bibinfo  {journal} {Phys. Rev. Lett.}\ }\textbf {\bibinfo {volume} {48}},\ \bibinfo {pages} {1559} (\bibinfo {year} {1982})}\BibitemShut {NoStop}%
\bibitem [{\citenamefont {Laughlin}(1983)}]{FQH2}%
  \BibitemOpen
  \bibfield  {author} {\bibinfo {author} {\bibfnamefont {R.~B.}\ \bibnamefont {Laughlin}},\ }\bibfield  {title} {\bibinfo {title} {Anomalous quantum hall effect: An incompressible quantum fluid with fractionally charged excitations},\ }\href {https://doi.org/10.1103/PhysRevLett.50.1395} {\bibfield  {journal} {\bibinfo  {journal} {Phys. Rev. Lett.}\ }\textbf {\bibinfo {volume} {50}},\ \bibinfo {pages} {1395} (\bibinfo {year} {1983})}\BibitemShut {NoStop}%
\bibitem [{\citenamefont {Stormer}\ \emph {et~al.}(1999)\citenamefont {Stormer}, \citenamefont {Tsui},\ and\ \citenamefont {Gossard}}]{FQH3}%
  \BibitemOpen
  \bibfield  {author} {\bibinfo {author} {\bibfnamefont {H.~L.}\ \bibnamefont {Stormer}}, \bibinfo {author} {\bibfnamefont {D.~C.}\ \bibnamefont {Tsui}},\ and\ \bibinfo {author} {\bibfnamefont {A.~C.}\ \bibnamefont {Gossard}},\ }\bibfield  {title} {\bibinfo {title} {The fractional quantum hall effect},\ }\href {https://doi.org/10.1103/RevModPhys.71.S298} {\bibfield  {journal} {\bibinfo  {journal} {Rev. Mod. Phys.}\ }\textbf {\bibinfo {volume} {71}},\ \bibinfo {pages} {S298} (\bibinfo {year} {1999})}\BibitemShut {NoStop}%
\bibitem [{\citenamefont {Haldane}(1983)}]{WF1}%
  \BibitemOpen
  \bibfield  {author} {\bibinfo {author} {\bibfnamefont {F.~D.~M.}\ \bibnamefont {Haldane}},\ }\bibfield  {title} {\bibinfo {title} {Fractional quantization of the hall effect: A hierarchy of incompressible quantum fluid states},\ }\href {https://doi.org/10.1103/PhysRevLett.51.605} {\bibfield  {journal} {\bibinfo  {journal} {Phys. Rev. Lett.}\ }\textbf {\bibinfo {volume} {51}},\ \bibinfo {pages} {605} (\bibinfo {year} {1983})}\BibitemShut {NoStop}%
\bibitem [{\citenamefont {Thouless}(1984)}]{WF2}%
  \BibitemOpen
  \bibfield  {author} {\bibinfo {author} {\bibfnamefont {D.}~\bibnamefont {Thouless}},\ }\bibfield  {title} {\bibinfo {title} {Theory of the quantized hall effect},\ }\href {https://doi.org/https://doi.org/10.1016/0039-6028(84)90299-1} {\bibfield  {journal} {\bibinfo  {journal} {Surface Science}\ }\textbf {\bibinfo {volume} {142}},\ \bibinfo {pages} {147} (\bibinfo {year} {1984})}\BibitemShut {NoStop}%
\bibitem [{\citenamefont {Haldane}\ and\ \citenamefont {Rezayi}(1985)}]{WF3}%
  \BibitemOpen
  \bibfield  {author} {\bibinfo {author} {\bibfnamefont {F.~D.~M.}\ \bibnamefont {Haldane}}\ and\ \bibinfo {author} {\bibfnamefont {E.~H.}\ \bibnamefont {Rezayi}},\ }\bibfield  {title} {\bibinfo {title} {Periodic laughlin-jastrow wave functions for the fractional quantized hall effect},\ }\href {https://doi.org/10.1103/PhysRevB.31.2529} {\bibfield  {journal} {\bibinfo  {journal} {Phys. Rev. B}\ }\textbf {\bibinfo {volume} {31}},\ \bibinfo {pages} {2529} (\bibinfo {year} {1985})}\BibitemShut {NoStop}%
\bibitem [{\citenamefont {Rezayi}\ and\ \citenamefont {Haldane}(1994)}]{WF4}%
  \BibitemOpen
  \bibfield  {author} {\bibinfo {author} {\bibfnamefont {E.~H.}\ \bibnamefont {Rezayi}}\ and\ \bibinfo {author} {\bibfnamefont {F.~D.~M.}\ \bibnamefont {Haldane}},\ }\bibfield  {title} {\bibinfo {title} {Laughlin state on stretched and squeezed cylinders and edge excitations in the quantum hall effect},\ }\href {https://doi.org/10.1103/PhysRevB.50.17199} {\bibfield  {journal} {\bibinfo  {journal} {Phys. Rev. B}\ }\textbf {\bibinfo {volume} {50}},\ \bibinfo {pages} {17199} (\bibinfo {year} {1994})}\BibitemShut {NoStop}%
\bibitem [{\citenamefont {Halperin}(1983)}]{Halperin1}%
  \BibitemOpen
  \bibfield  {author} {\bibinfo {author} {\bibfnamefont {B.~I.}\ \bibnamefont {Halperin}},\ }\bibfield  {title} {\bibinfo {title} {Theory of the quantized hall conductance},\ }\href {https://api.semanticscholar.org/CorpusID:117562779} {\bibfield  {journal} {\bibinfo  {journal} {Helvetica Physica Acta}\ }\textbf {\bibinfo {volume} {56}},\ \bibinfo {pages} {75} (\bibinfo {year} {1983})}\BibitemShut {NoStop}%
\bibitem [{\citenamefont {Halperin}(1984)}]{Halperin2}%
  \BibitemOpen
  \bibfield  {author} {\bibinfo {author} {\bibfnamefont {B.~I.}\ \bibnamefont {Halperin}},\ }\bibfield  {title} {\bibinfo {title} {Statistics of quasiparticles and the hierarchy of fractional quantized hall states},\ }\href {https://doi.org/10.1103/PhysRevLett.52.1583} {\bibfield  {journal} {\bibinfo  {journal} {Phys. Rev. Lett.}\ }\textbf {\bibinfo {volume} {52}},\ \bibinfo {pages} {1583} (\bibinfo {year} {1984})}\BibitemShut {NoStop}%
\bibitem [{\citenamefont {Jain}(1989)}]{WF5}%
  \BibitemOpen
  \bibfield  {author} {\bibinfo {author} {\bibfnamefont {J.~K.}\ \bibnamefont {Jain}},\ }\bibfield  {title} {\bibinfo {title} {Composite-fermion approach for the fractional quantum hall effect},\ }\href {https://doi.org/10.1103/PhysRevLett.63.199} {\bibfield  {journal} {\bibinfo  {journal} {Phys. Rev. Lett.}\ }\textbf {\bibinfo {volume} {63}},\ \bibinfo {pages} {199} (\bibinfo {year} {1989})}\BibitemShut {NoStop}%
\bibitem [{\citenamefont {Bernevig}\ and\ \citenamefont {Haldane}(2008{\natexlab{a}})}]{Jacks1}%
  \BibitemOpen
  \bibfield  {author} {\bibinfo {author} {\bibfnamefont {B.~A.}\ \bibnamefont {Bernevig}}\ and\ \bibinfo {author} {\bibfnamefont {F.~D.~M.}\ \bibnamefont {Haldane}},\ }\bibfield  {title} {\bibinfo {title} {Model fractional quantum hall states and jack polynomials},\ }\href {https://doi.org/10.1103/PhysRevLett.100.246802} {\bibfield  {journal} {\bibinfo  {journal} {Phys. Rev. Lett.}\ }\textbf {\bibinfo {volume} {100}},\ \bibinfo {pages} {246802} (\bibinfo {year} {2008}{\natexlab{a}})}\BibitemShut {NoStop}%
\bibitem [{\citenamefont {Bernevig}\ and\ \citenamefont {Haldane}(2008{\natexlab{b}})}]{Jacks2}%
  \BibitemOpen
  \bibfield  {author} {\bibinfo {author} {\bibfnamefont {B.~A.}\ \bibnamefont {Bernevig}}\ and\ \bibinfo {author} {\bibfnamefont {F.~D.~M.}\ \bibnamefont {Haldane}},\ }\bibfield  {title} {\bibinfo {title} {Generalized clustering conditions of jack polynomials at negative jack parameter $\ensuremath{\alpha}$},\ }\href {https://doi.org/10.1103/PhysRevB.77.184502} {\bibfield  {journal} {\bibinfo  {journal} {Phys. Rev. B}\ }\textbf {\bibinfo {volume} {77}},\ \bibinfo {pages} {184502} (\bibinfo {year} {2008}{\natexlab{b}})}\BibitemShut {NoStop}%
\bibitem [{\citenamefont {Bernevig}\ and\ \citenamefont {Haldane}(2008{\natexlab{c}})}]{Jacks3}%
  \BibitemOpen
  \bibfield  {author} {\bibinfo {author} {\bibfnamefont {B.~A.}\ \bibnamefont {Bernevig}}\ and\ \bibinfo {author} {\bibfnamefont {F.~D.~M.}\ \bibnamefont {Haldane}},\ }\bibfield  {title} {\bibinfo {title} {Properties of non-abelian fractional quantum hall states at filling $\ensuremath{\nu}=k/r$},\ }\href {https://doi.org/10.1103/PhysRevLett.101.246806} {\bibfield  {journal} {\bibinfo  {journal} {Phys. Rev. Lett.}\ }\textbf {\bibinfo {volume} {101}},\ \bibinfo {pages} {246806} (\bibinfo {year} {2008}{\natexlab{c}})}\BibitemShut {NoStop}%
\bibitem [{\citenamefont {Bernevig}\ and\ \citenamefont {Haldane}(2009)}]{Jacks4}%
  \BibitemOpen
  \bibfield  {author} {\bibinfo {author} {\bibfnamefont {B.~A.}\ \bibnamefont {Bernevig}}\ and\ \bibinfo {author} {\bibfnamefont {F.~D.~M.}\ \bibnamefont {Haldane}},\ }\bibfield  {title} {\bibinfo {title} {Clustering properties and model wave functions for non-abelian fractional quantum hall quasielectrons},\ }\href {https://doi.org/10.1103/PhysRevLett.102.066802} {\bibfield  {journal} {\bibinfo  {journal} {Phys. Rev. Lett.}\ }\textbf {\bibinfo {volume} {102}},\ \bibinfo {pages} {066802} (\bibinfo {year} {2009})}\BibitemShut {NoStop}%
\bibitem [{\citenamefont {Bernevig}\ and\ \citenamefont {Regnault}(2009)}]{Jacks5}%
  \BibitemOpen
  \bibfield  {author} {\bibinfo {author} {\bibfnamefont {B.~A.}\ \bibnamefont {Bernevig}}\ and\ \bibinfo {author} {\bibfnamefont {N.}~\bibnamefont {Regnault}},\ }\bibfield  {title} {\bibinfo {title} {Anatomy of abelian and non-abelian fractional quantum hall states},\ }\href {https://doi.org/10.1103/PhysRevLett.103.206801} {\bibfield  {journal} {\bibinfo  {journal} {Phys. Rev. Lett.}\ }\textbf {\bibinfo {volume} {103}},\ \bibinfo {pages} {206801} (\bibinfo {year} {2009})}\BibitemShut {NoStop}%
\bibitem [{\citenamefont {Thomale}\ \emph {et~al.}(2011)\citenamefont {Thomale}, \citenamefont {Estienne}, \citenamefont {Regnault},\ and\ \citenamefont {Bernevig}}]{Jacks6}%
  \BibitemOpen
  \bibfield  {author} {\bibinfo {author} {\bibfnamefont {R.}~\bibnamefont {Thomale}}, \bibinfo {author} {\bibfnamefont {B.}~\bibnamefont {Estienne}}, \bibinfo {author} {\bibfnamefont {N.}~\bibnamefont {Regnault}},\ and\ \bibinfo {author} {\bibfnamefont {B.~A.}\ \bibnamefont {Bernevig}},\ }\bibfield  {title} {\bibinfo {title} {Decomposition of fractional quantum hall model states: Product rule symmetries and approximations},\ }\href {https://doi.org/10.1103/PhysRevB.84.045127} {\bibfield  {journal} {\bibinfo  {journal} {Phys. Rev. B}\ }\textbf {\bibinfo {volume} {84}},\ \bibinfo {pages} {045127} (\bibinfo {year} {2011})}\BibitemShut {NoStop}%
\bibitem [{\citenamefont {Ardonne}\ and\ \citenamefont {Regnault}(2011)}]{Jacks7}%
  \BibitemOpen
  \bibfield  {author} {\bibinfo {author} {\bibfnamefont {E.}~\bibnamefont {Ardonne}}\ and\ \bibinfo {author} {\bibfnamefont {N.}~\bibnamefont {Regnault}},\ }\bibfield  {title} {\bibinfo {title} {Structure of spinful quantum hall states: A squeezing perspective},\ }\href {https://doi.org/10.1103/PhysRevB.84.205134} {\bibfield  {journal} {\bibinfo  {journal} {Phys. Rev. B}\ }\textbf {\bibinfo {volume} {84}},\ \bibinfo {pages} {205134} (\bibinfo {year} {2011})}\BibitemShut {NoStop}%
\bibitem [{\citenamefont {Estienne}\ \emph {et~al.}(2010)\citenamefont {Estienne}, \citenamefont {Regnault},\ and\ \citenamefont {Santachiara}}]{Jacks8}%
  \BibitemOpen
  \bibfield  {author} {\bibinfo {author} {\bibfnamefont {B.}~\bibnamefont {Estienne}}, \bibinfo {author} {\bibfnamefont {N.}~\bibnamefont {Regnault}},\ and\ \bibinfo {author} {\bibfnamefont {R.}~\bibnamefont {Santachiara}},\ }\bibfield  {title} {\bibinfo {title} {Clustering properties, jack polynomials and unitary conformal field theories},\ }\href {https://doi.org/https://doi.org/10.1016/j.nuclphysb.2009.09.002} {\bibfield  {journal} {\bibinfo  {journal} {Nuclear Physics B}\ }\textbf {\bibinfo {volume} {824}},\ \bibinfo {pages} {539} (\bibinfo {year} {2010})}\BibitemShut {NoStop}%
\bibitem [{\citenamefont {Estienne}\ and\ \citenamefont {Bernevig}(2012)}]{Jacks9}%
  \BibitemOpen
  \bibfield  {author} {\bibinfo {author} {\bibfnamefont {B.}~\bibnamefont {Estienne}}\ and\ \bibinfo {author} {\bibfnamefont {B.~A.}\ \bibnamefont {Bernevig}},\ }\bibfield  {title} {\bibinfo {title} {Spin-singlet quantum hall states and jack polynomials with a prescribed symmetry},\ }\href {https://doi.org/https://doi.org/10.1016/j.nuclphysb.2011.12.007} {\bibfield  {journal} {\bibinfo  {journal} {Nuclear Physics B}\ }\textbf {\bibinfo {volume} {857}},\ \bibinfo {pages} {185} (\bibinfo {year} {2012})}\BibitemShut {NoStop}%
\bibitem [{\citenamefont {Haldane}(1991)}]{GPP1}%
  \BibitemOpen
  \bibfield  {author} {\bibinfo {author} {\bibfnamefont {F.~D.~M.}\ \bibnamefont {Haldane}},\ }\bibfield  {title} {\bibinfo {title} {``fractional statistics'' in arbitrary dimensions: A generalization of the pauli principle},\ }\href {https://doi.org/10.1103/PhysRevLett.67.937} {\bibfield  {journal} {\bibinfo  {journal} {Phys. Rev. Lett.}\ }\textbf {\bibinfo {volume} {67}},\ \bibinfo {pages} {937} (\bibinfo {year} {1991})}\BibitemShut {NoStop}%
\bibitem [{\citenamefont {Wu}(1994)}]{GPP2}%
  \BibitemOpen
  \bibfield  {author} {\bibinfo {author} {\bibfnamefont {Y.-S.}\ \bibnamefont {Wu}},\ }\bibfield  {title} {\bibinfo {title} {Statistical distribution for generalized ideal gas of fractional-statistics particles},\ }\href {https://doi.org/10.1103/PhysRevLett.73.922} {\bibfield  {journal} {\bibinfo  {journal} {Phys. Rev. Lett.}\ }\textbf {\bibinfo {volume} {73}},\ \bibinfo {pages} {922} (\bibinfo {year} {1994})}\BibitemShut {NoStop}%
\bibitem [{SM()}]{SM}%
  \BibitemOpen
  \href@noop {} {\ }\bibinfo {note} {See Supplemental Material for details on the decomposition of other Laughlin states, Halperin (3,3,2) state, the corresponding squeezing operation, basis dimension of FQH state, orbital entanglement spectrum and many-body density profile.}\BibitemShut {Stop}%
\bibitem [{\citenamefont {Li}\ and\ \citenamefont {Haldane}(2008)}]{LiHaldane}%
  \BibitemOpen
  \bibfield  {author} {\bibinfo {author} {\bibfnamefont {H.}~\bibnamefont {Li}}\ and\ \bibinfo {author} {\bibfnamefont {F.~D.~M.}\ \bibnamefont {Haldane}},\ }\bibfield  {title} {\bibinfo {title} {Entanglement spectrum as a generalization of entanglement entropy: Identification of topological order in non-abelian fractional quantum hall effect states},\ }\href {https://doi.org/10.1103/PhysRevLett.101.010504} {\bibfield  {journal} {\bibinfo  {journal} {Phys. Rev. Lett.}\ }\textbf {\bibinfo {volume} {101}},\ \bibinfo {pages} {010504} (\bibinfo {year} {2008})}\BibitemShut {NoStop}%
\bibitem [{\citenamefont {Wen}(1995)}]{XGWen}%
  \BibitemOpen
  \bibfield  {author} {\bibinfo {author} {\bibfnamefont {X.-G.}\ \bibnamefont {Wen}},\ }\bibfield  {title} {\bibinfo {title} {Topological orders and edge excitations in fractional quantum hall states},\ }\href {https://doi.org/10.1080/00018739500101566} {\bibfield  {journal} {\bibinfo  {journal} {Advances in Physics}\ }\textbf {\bibinfo {volume} {44}},\ \bibinfo {pages} {405} (\bibinfo {year} {1995})},\ \Eprint {https://arxiv.org/abs/https://doi.org/10.1080/00018739500101566} {https://doi.org/10.1080/00018739500101566} \BibitemShut {NoStop}%
\bibitem [{\citenamefont {He}\ \emph {et~al.}(2015)\citenamefont {He}, \citenamefont {Luo}, \citenamefont {Wang},\ and\ \citenamefont {Gong}}]{He_2015}%
  \BibitemOpen
  \bibfield  {author} {\bibinfo {author} {\bibfnamefont {A.-L.}\ \bibnamefont {He}}, \bibinfo {author} {\bibfnamefont {W.-W.}\ \bibnamefont {Luo}}, \bibinfo {author} {\bibfnamefont {Y.-F.}\ \bibnamefont {Wang}},\ and\ \bibinfo {author} {\bibfnamefont {C.-D.}\ \bibnamefont {Gong}},\ }\bibfield  {title} {\bibinfo {title} {Wave functions for fractional chern insulators in disk geometry},\ }\href {https://doi.org/10.1088/1367-2630/17/12/125005} {\bibfield  {journal} {\bibinfo  {journal} {New Journal of Physics}\ }\textbf {\bibinfo {volume} {17}},\ \bibinfo {pages} {125005} (\bibinfo {year} {2015})}\BibitemShut {NoStop}%
\bibitem [{\citenamefont {He}\ \emph {et~al.}(2020)\citenamefont {He}, \citenamefont {Luo}, \citenamefont {Yao},\ and\ \citenamefont {Wang}}]{NAFCIs1}%
  \BibitemOpen
  \bibfield  {author} {\bibinfo {author} {\bibfnamefont {A.-L.}\ \bibnamefont {He}}, \bibinfo {author} {\bibfnamefont {W.-W.}\ \bibnamefont {Luo}}, \bibinfo {author} {\bibfnamefont {H.}~\bibnamefont {Yao}},\ and\ \bibinfo {author} {\bibfnamefont {Y.-F.}\ \bibnamefont {Wang}},\ }\bibfield  {title} {\bibinfo {title} {Non-abelian fractional chern insulator in disk geometry},\ }\href {https://doi.org/10.1103/PhysRevB.101.165127} {\bibfield  {journal} {\bibinfo  {journal} {Phys. Rev. B}\ }\textbf {\bibinfo {volume} {101}},\ \bibinfo {pages} {165127} (\bibinfo {year} {2020})}\BibitemShut {NoStop}%
\bibitem [{\citenamefont {Wu}\ \emph {et~al.}(2013)\citenamefont {Wu}, \citenamefont {Regnault},\ and\ \citenamefont {Bernevig}}]{colorent1}%
  \BibitemOpen
  \bibfield  {author} {\bibinfo {author} {\bibfnamefont {Y.-L.}\ \bibnamefont {Wu}}, \bibinfo {author} {\bibfnamefont {N.}~\bibnamefont {Regnault}},\ and\ \bibinfo {author} {\bibfnamefont {B.~A.}\ \bibnamefont {Bernevig}},\ }\bibfield  {title} {\bibinfo {title} {Bloch model wave functions and pseudopotentials for all fractional chern insulators},\ }\href {https://doi.org/10.1103/PhysRevLett.110.106802} {\bibfield  {journal} {\bibinfo  {journal} {Phys. Rev. Lett.}\ }\textbf {\bibinfo {volume} {110}},\ \bibinfo {pages} {106802} (\bibinfo {year} {2013})}\BibitemShut {NoStop}%
\bibitem [{\citenamefont {Wang}\ \emph {et~al.}(2023)\citenamefont {Wang}, \citenamefont {Klevtsov},\ and\ \citenamefont {Liu}}]{colorent2}%
  \BibitemOpen
  \bibfield  {author} {\bibinfo {author} {\bibfnamefont {J.}~\bibnamefont {Wang}}, \bibinfo {author} {\bibfnamefont {S.}~\bibnamefont {Klevtsov}},\ and\ \bibinfo {author} {\bibfnamefont {Z.}~\bibnamefont {Liu}},\ }\bibfield  {title} {\bibinfo {title} {Origin of model fractional chern insulators in all topological ideal flatbands: Explicit color-entangled wave function and exact density algebra},\ }\href {https://doi.org/10.1103/PhysRevResearch.5.023167} {\bibfield  {journal} {\bibinfo  {journal} {Phys. Rev. Res.}\ }\textbf {\bibinfo {volume} {5}},\ \bibinfo {pages} {023167} (\bibinfo {year} {2023})}\BibitemShut {NoStop}%
\bibitem [{\citenamefont {Zeng}\ \emph {et~al.}(2017)\citenamefont {Zeng}, \citenamefont {Zhu},\ and\ \citenamefont {Sheng}}]{zeng1}%
  \BibitemOpen
  \bibfield  {author} {\bibinfo {author} {\bibfnamefont {T.-S.}\ \bibnamefont {Zeng}}, \bibinfo {author} {\bibfnamefont {W.}~\bibnamefont {Zhu}},\ and\ \bibinfo {author} {\bibfnamefont {D.~N.}\ \bibnamefont {Sheng}},\ }\bibfield  {title} {\bibinfo {title} {Two-component quantum hall effects in topological flat bands},\ }\href {https://doi.org/10.1103/PhysRevB.95.125134} {\bibfield  {journal} {\bibinfo  {journal} {Phys. Rev. B}\ }\textbf {\bibinfo {volume} {95}},\ \bibinfo {pages} {125134} (\bibinfo {year} {2017})}\BibitemShut {NoStop}%
\bibitem [{\citenamefont {Zeng}\ and\ \citenamefont {Sheng}(2018)}]{zeng2}%
  \BibitemOpen
  \bibfield  {author} {\bibinfo {author} {\bibfnamefont {T.-S.}\ \bibnamefont {Zeng}}\ and\ \bibinfo {author} {\bibfnamefont {D.~N.}\ \bibnamefont {Sheng}},\ }\bibfield  {title} {\bibinfo {title} {$\mathrm{SU}(n)$ fractional quantum hall effect in topological flat bands},\ }\href {https://doi.org/10.1103/PhysRevB.97.035151} {\bibfield  {journal} {\bibinfo  {journal} {Phys. Rev. B}\ }\textbf {\bibinfo {volume} {97}},\ \bibinfo {pages} {035151} (\bibinfo {year} {2018})}\BibitemShut {NoStop}%
\bibitem [{\citenamefont {Zeng}\ \emph {et~al.}(2019)\citenamefont {Zeng}, \citenamefont {Sheng},\ and\ \citenamefont {Zhu}}]{zeng21}%
  \BibitemOpen
  \bibfield  {author} {\bibinfo {author} {\bibfnamefont {T.-S.}\ \bibnamefont {Zeng}}, \bibinfo {author} {\bibfnamefont {D.~N.}\ \bibnamefont {Sheng}},\ and\ \bibinfo {author} {\bibfnamefont {W.}~\bibnamefont {Zhu}},\ }\bibfield  {title} {\bibinfo {title} {Topological characterization of hierarchical fractional quantum hall effects in topological flat bands with su($n$) symmetry},\ }\href {https://doi.org/10.1103/PhysRevB.100.075106} {\bibfield  {journal} {\bibinfo  {journal} {Phys. Rev. B}\ }\textbf {\bibinfo {volume} {100}},\ \bibinfo {pages} {075106} (\bibinfo {year} {2019})}\BibitemShut {NoStop}%
\bibitem [{\citenamefont {Zeng}\ and\ \citenamefont {Zhu}(2022)}]{zeng3}%
  \BibitemOpen
  \bibfield  {author} {\bibinfo {author} {\bibfnamefont {T.-S.}\ \bibnamefont {Zeng}}\ and\ \bibinfo {author} {\bibfnamefont {W.}~\bibnamefont {Zhu}},\ }\bibfield  {title} {\bibinfo {title} {Chern-number matrix of the non-abelian spin-singlet fractional quantum hall effect},\ }\href {https://doi.org/10.1103/PhysRevB.105.125128} {\bibfield  {journal} {\bibinfo  {journal} {Phys. Rev. B}\ }\textbf {\bibinfo {volume} {105}},\ \bibinfo {pages} {125128} (\bibinfo {year} {2022})}\BibitemShut {NoStop}%
\bibitem [{\citenamefont {Cai}\ \emph {et~al.}(2023)\citenamefont {Cai}, \citenamefont {Anderson}, \citenamefont {Wang}, \citenamefont {Zhang}, \citenamefont {Liu}, \citenamefont {Holtzmann}, \citenamefont {Zhang}, \citenamefont {Fan}, \citenamefont {Taniguchi}, \citenamefont {Watanabe}, \citenamefont {Ran}, \citenamefont {Cao}, \citenamefont {Fu}, \citenamefont {Xiao}, \citenamefont {Yao},\ and\ \citenamefont {Xu}}]{ReaFCI1}%
  \BibitemOpen
  \bibfield  {author} {\bibinfo {author} {\bibfnamefont {J.}~\bibnamefont {Cai}}, \bibinfo {author} {\bibfnamefont {E.}~\bibnamefont {Anderson}}, \bibinfo {author} {\bibfnamefont {C.}~\bibnamefont {Wang}}, \bibinfo {author} {\bibfnamefont {X.}~\bibnamefont {Zhang}}, \bibinfo {author} {\bibfnamefont {X.}~\bibnamefont {Liu}}, \bibinfo {author} {\bibfnamefont {W.}~\bibnamefont {Holtzmann}}, \bibinfo {author} {\bibfnamefont {Y.}~\bibnamefont {Zhang}}, \bibinfo {author} {\bibfnamefont {F.}~\bibnamefont {Fan}}, \bibinfo {author} {\bibfnamefont {T.}~\bibnamefont {Taniguchi}}, \bibinfo {author} {\bibfnamefont {K.}~\bibnamefont {Watanabe}}, \bibinfo {author} {\bibfnamefont {Y.}~\bibnamefont {Ran}}, \bibinfo {author} {\bibfnamefont {T.}~\bibnamefont {Cao}}, \bibinfo {author} {\bibfnamefont {L.}~\bibnamefont {Fu}}, \bibinfo {author} {\bibfnamefont {D.}~\bibnamefont {Xiao}}, \bibinfo {author} {\bibfnamefont {W.}~\bibnamefont {Yao}},\ and\ \bibinfo {author} {\bibfnamefont {X.}~\bibnamefont {Xu}},\ }\bibfield  {title}
  {\bibinfo {title} {Signatures of fractional quantum anomalous hall states in twisted mote2},\ }\href {https://doi.org/10.1038/s41586-023-06289-w} {\bibfield  {journal} {\bibinfo  {journal} {Nature}\ }\textbf {\bibinfo {volume} {622}},\ \bibinfo {pages} {63} (\bibinfo {year} {2023})}\BibitemShut {NoStop}%
\bibitem [{\citenamefont {Zeng}\ \emph {et~al.}(2023)\citenamefont {Zeng}, \citenamefont {Xia}, \citenamefont {Kang}, \citenamefont {Zhu}, \citenamefont {Kn{\"u}ppel}, \citenamefont {Vaswani}, \citenamefont {Watanabe}, \citenamefont {Taniguchi}, \citenamefont {Mak},\ and\ \citenamefont {Shan}}]{ReaFCI2}%
  \BibitemOpen
  \bibfield  {author} {\bibinfo {author} {\bibfnamefont {Y.}~\bibnamefont {Zeng}}, \bibinfo {author} {\bibfnamefont {Z.}~\bibnamefont {Xia}}, \bibinfo {author} {\bibfnamefont {K.}~\bibnamefont {Kang}}, \bibinfo {author} {\bibfnamefont {J.}~\bibnamefont {Zhu}}, \bibinfo {author} {\bibfnamefont {P.}~\bibnamefont {Kn{\"u}ppel}}, \bibinfo {author} {\bibfnamefont {C.}~\bibnamefont {Vaswani}}, \bibinfo {author} {\bibfnamefont {K.}~\bibnamefont {Watanabe}}, \bibinfo {author} {\bibfnamefont {T.}~\bibnamefont {Taniguchi}}, \bibinfo {author} {\bibfnamefont {K.~F.}\ \bibnamefont {Mak}},\ and\ \bibinfo {author} {\bibfnamefont {J.}~\bibnamefont {Shan}},\ }\bibfield  {title} {\bibinfo {title} {Thermodynamic evidence of fractional chern insulator in moir{\'e} mote2},\ }\href {https://doi.org/10.1038/s41586-023-06452-3} {\bibfield  {journal} {\bibinfo  {journal} {Nature}\ }\textbf {\bibinfo {volume} {622}},\ \bibinfo {pages} {69} (\bibinfo {year} {2023})}\BibitemShut {NoStop}%
\bibitem [{\citenamefont {Park}\ \emph {et~al.}(2023)\citenamefont {Park}, \citenamefont {Cai}, \citenamefont {Anderson}, \citenamefont {Zhang}, \citenamefont {Zhu}, \citenamefont {Liu}, \citenamefont {Wang}, \citenamefont {Holtzmann}, \citenamefont {Hu}, \citenamefont {Liu}, \citenamefont {Taniguchi}, \citenamefont {Watanabe}, \citenamefont {Chu}, \citenamefont {Cao}, \citenamefont {Fu}, \citenamefont {Yao}, \citenamefont {Chang}, \citenamefont {Cobden}, \citenamefont {Xiao},\ and\ \citenamefont {Xu}}]{ReaFCI3}%
  \BibitemOpen
  \bibfield  {author} {\bibinfo {author} {\bibfnamefont {H.}~\bibnamefont {Park}}, \bibinfo {author} {\bibfnamefont {J.}~\bibnamefont {Cai}}, \bibinfo {author} {\bibfnamefont {E.}~\bibnamefont {Anderson}}, \bibinfo {author} {\bibfnamefont {Y.}~\bibnamefont {Zhang}}, \bibinfo {author} {\bibfnamefont {J.}~\bibnamefont {Zhu}}, \bibinfo {author} {\bibfnamefont {X.}~\bibnamefont {Liu}}, \bibinfo {author} {\bibfnamefont {C.}~\bibnamefont {Wang}}, \bibinfo {author} {\bibfnamefont {W.}~\bibnamefont {Holtzmann}}, \bibinfo {author} {\bibfnamefont {C.}~\bibnamefont {Hu}}, \bibinfo {author} {\bibfnamefont {Z.}~\bibnamefont {Liu}}, \bibinfo {author} {\bibfnamefont {T.}~\bibnamefont {Taniguchi}}, \bibinfo {author} {\bibfnamefont {K.}~\bibnamefont {Watanabe}}, \bibinfo {author} {\bibfnamefont {J.-H.}\ \bibnamefont {Chu}}, \bibinfo {author} {\bibfnamefont {T.}~\bibnamefont {Cao}}, \bibinfo {author} {\bibfnamefont {L.}~\bibnamefont {Fu}}, \bibinfo {author} {\bibfnamefont {W.}~\bibnamefont {Yao}}, \bibinfo {author}
  {\bibfnamefont {C.-Z.}\ \bibnamefont {Chang}}, \bibinfo {author} {\bibfnamefont {D.}~\bibnamefont {Cobden}}, \bibinfo {author} {\bibfnamefont {D.}~\bibnamefont {Xiao}},\ and\ \bibinfo {author} {\bibfnamefont {X.}~\bibnamefont {Xu}},\ }\bibfield  {title} {\bibinfo {title} {Observation of fractionally quantized anomalous hall effect},\ }\href {https://doi.org/10.1038/s41586-023-06536-0} {\bibfield  {journal} {\bibinfo  {journal} {Nature}\ }\textbf {\bibinfo {volume} {622}},\ \bibinfo {pages} {74} (\bibinfo {year} {2023})}\BibitemShut {NoStop}%
\bibitem [{\citenamefont {Xu}\ \emph {et~al.}(2023)\citenamefont {Xu}, \citenamefont {Sun}, \citenamefont {Jia}, \citenamefont {Liu}, \citenamefont {Xu}, \citenamefont {Li}, \citenamefont {Gu}, \citenamefont {Watanabe}, \citenamefont {Taniguchi}, \citenamefont {Tong}, \citenamefont {Jia}, \citenamefont {Shi}, \citenamefont {Jiang}, \citenamefont {Zhang}, \citenamefont {Liu},\ and\ \citenamefont {Li}}]{ReaFCI4}%
  \BibitemOpen
  \bibfield  {author} {\bibinfo {author} {\bibfnamefont {F.}~\bibnamefont {Xu}}, \bibinfo {author} {\bibfnamefont {Z.}~\bibnamefont {Sun}}, \bibinfo {author} {\bibfnamefont {T.}~\bibnamefont {Jia}}, \bibinfo {author} {\bibfnamefont {C.}~\bibnamefont {Liu}}, \bibinfo {author} {\bibfnamefont {C.}~\bibnamefont {Xu}}, \bibinfo {author} {\bibfnamefont {C.}~\bibnamefont {Li}}, \bibinfo {author} {\bibfnamefont {Y.}~\bibnamefont {Gu}}, \bibinfo {author} {\bibfnamefont {K.}~\bibnamefont {Watanabe}}, \bibinfo {author} {\bibfnamefont {T.}~\bibnamefont {Taniguchi}}, \bibinfo {author} {\bibfnamefont {B.}~\bibnamefont {Tong}}, \bibinfo {author} {\bibfnamefont {J.}~\bibnamefont {Jia}}, \bibinfo {author} {\bibfnamefont {Z.}~\bibnamefont {Shi}}, \bibinfo {author} {\bibfnamefont {S.}~\bibnamefont {Jiang}}, \bibinfo {author} {\bibfnamefont {Y.}~\bibnamefont {Zhang}}, \bibinfo {author} {\bibfnamefont {X.}~\bibnamefont {Liu}},\ and\ \bibinfo {author} {\bibfnamefont {T.}~\bibnamefont {Li}},\ }\bibfield  {title} {\bibinfo {title}
  {Observation of integer and fractional quantum anomalous hall effects in twisted bilayer ${\mathrm{mote}}_{2}$},\ }\href {https://doi.org/10.1103/PhysRevX.13.031037} {\bibfield  {journal} {\bibinfo  {journal} {Phys. Rev. X}\ }\textbf {\bibinfo {volume} {13}},\ \bibinfo {pages} {031037} (\bibinfo {year} {2023})}\BibitemShut {NoStop}%
\bibitem [{\citenamefont {Lu}\ \emph {et~al.}(2024)\citenamefont {Lu}, \citenamefont {Han}, \citenamefont {Yao}, \citenamefont {Reddy}, \citenamefont {Yang}, \citenamefont {Seo}, \citenamefont {Watanabe}, \citenamefont {Taniguchi}, \citenamefont {Fu},\ and\ \citenamefont {Ju}}]{ReaFCI5}%
  \BibitemOpen
  \bibfield  {author} {\bibinfo {author} {\bibfnamefont {Z.}~\bibnamefont {Lu}}, \bibinfo {author} {\bibfnamefont {T.}~\bibnamefont {Han}}, \bibinfo {author} {\bibfnamefont {Y.}~\bibnamefont {Yao}}, \bibinfo {author} {\bibfnamefont {A.~P.}\ \bibnamefont {Reddy}}, \bibinfo {author} {\bibfnamefont {J.}~\bibnamefont {Yang}}, \bibinfo {author} {\bibfnamefont {J.}~\bibnamefont {Seo}}, \bibinfo {author} {\bibfnamefont {K.}~\bibnamefont {Watanabe}}, \bibinfo {author} {\bibfnamefont {T.}~\bibnamefont {Taniguchi}}, \bibinfo {author} {\bibfnamefont {L.}~\bibnamefont {Fu}},\ and\ \bibinfo {author} {\bibfnamefont {L.}~\bibnamefont {Ju}},\ }\bibfield  {title} {\bibinfo {title} {Fractional quantum anomalous hall effect in multilayer graphene},\ }\href {https://doi.org/10.1038/s41586-023-07010-7} {\bibfield  {journal} {\bibinfo  {journal} {Nature}\ }\textbf {\bibinfo {volume} {626}},\ \bibinfo {pages} {759} (\bibinfo {year} {2024})}\BibitemShut {NoStop}%
\bibitem [{\citenamefont {Dong}\ \emph {et~al.}(2025)\citenamefont {Dong}, \citenamefont {Liu}, \citenamefont {Zhu}, \citenamefont {Pan}, \citenamefont {Hong}, \citenamefont {Wang}, \citenamefont {Ren}, \citenamefont {Jia}, \citenamefont {Watanabe}, \citenamefont {Taniguchi}, \citenamefont {Du}, \citenamefont {Shi}, \citenamefont {Yang},\ and\ \citenamefont {Zhang}}]{ReaFCI6}%
  \BibitemOpen
  \bibfield  {author} {\bibinfo {author} {\bibfnamefont {J.}~\bibnamefont {Dong}}, \bibinfo {author} {\bibfnamefont {L.}~\bibnamefont {Liu}}, \bibinfo {author} {\bibfnamefont {J.}~\bibnamefont {Zhu}}, \bibinfo {author} {\bibfnamefont {Z.}~\bibnamefont {Pan}}, \bibinfo {author} {\bibfnamefont {Y.}~\bibnamefont {Hong}}, \bibinfo {author} {\bibfnamefont {F.}~\bibnamefont {Wang}}, \bibinfo {author} {\bibfnamefont {Z.}~\bibnamefont {Ren}}, \bibinfo {author} {\bibfnamefont {Z.}~\bibnamefont {Jia}}, \bibinfo {author} {\bibfnamefont {K.}~\bibnamefont {Watanabe}}, \bibinfo {author} {\bibfnamefont {T.}~\bibnamefont {Taniguchi}}, \bibinfo {author} {\bibfnamefont {L.}~\bibnamefont {Du}}, \bibinfo {author} {\bibfnamefont {D.}~\bibnamefont {Shi}}, \bibinfo {author} {\bibfnamefont {W.}~\bibnamefont {Yang}},\ and\ \bibinfo {author} {\bibfnamefont {G.}~\bibnamefont {Zhang}},\ }\href {https://arxiv.org/abs/2507.09908} {\bibinfo {title} {Observation of integer and fractional chern insulators in high chern number flatbands}}
  (\bibinfo {year} {2025}),\ \Eprint {https://arxiv.org/abs/2507.09908} {arXiv:2507.09908 [cond-mat.str-el]} \BibitemShut {NoStop}%
\bibitem [{\citenamefont {Li}\ \emph {et~al.}(2025)\citenamefont {Li}, \citenamefont {Wang}, \citenamefont {Wang}, \citenamefont {Zhang}, \citenamefont {Yang}, \citenamefont {Watanabe}, \citenamefont {Taniguchi}, \citenamefont {Xie}, \citenamefont {Wang}, \citenamefont {Liu}, \citenamefont {Song},\ and\ \citenamefont {Lu}}]{ReaFCI7}%
  \BibitemOpen
  \bibfield  {author} {\bibinfo {author} {\bibfnamefont {Z.}~\bibnamefont {Li}}, \bibinfo {author} {\bibfnamefont {W.}~\bibnamefont {Wang}}, \bibinfo {author} {\bibfnamefont {F.}~\bibnamefont {Wang}}, \bibinfo {author} {\bibfnamefont {Z.}~\bibnamefont {Zhang}}, \bibinfo {author} {\bibfnamefont {Q.}~\bibnamefont {Yang}}, \bibinfo {author} {\bibfnamefont {K.}~\bibnamefont {Watanabe}}, \bibinfo {author} {\bibfnamefont {T.}~\bibnamefont {Taniguchi}}, \bibinfo {author} {\bibfnamefont {X.~C.}\ \bibnamefont {Xie}}, \bibinfo {author} {\bibfnamefont {J.}~\bibnamefont {Wang}}, \bibinfo {author} {\bibfnamefont {K.}~\bibnamefont {Liu}}, \bibinfo {author} {\bibfnamefont {Z.}~\bibnamefont {Song}},\ and\ \bibinfo {author} {\bibfnamefont {X.}~\bibnamefont {Lu}},\ }\href {https://arxiv.org/abs/2512.21612} {\bibinfo {title} {Fractionalization and entanglement of high chern insulators}} (\bibinfo {year} {2025}),\ \Eprint {https://arxiv.org/abs/2512.21612} {arXiv:2512.21612 [cond-mat.mes-hall]} \BibitemShut {NoStop}%
\end{thebibliography}%
\end{document}